# Viscoelastic properties and flow instabilities of aqueous suspensions of cellulosic fibers

Jing He[1,2], Stephanie S. Lee[2], Manuk Colakyan[4] and Dilhan M. Kalyon[1-3, *]

Highly Filled Materials Institute[1]

Chemical Engineering and Materials Science[2]

Biomedical Engineering[3]

Stevens Institute of Technology

1 Castle Point Terrace, Hoboken, NJ 07030, USA

Particulate Solid Research[4], Inc. Chicago, IL 60632

\* For correspondence, dkalyon@stevens.edu

**Abstract**

Processing of concentrated lignocellulosic biomass suspensions typically involves the conversion of the cellulose into sugars and sugars into ethanol. Biomass is usually pre-processed (i.e. via comminution, steam explosion, etc.) to form fine cellulosic fibers to be dispersed into an aqueous phase for further treatment. The resulting cellulose suspensions need to be pressurized and pumped into and out of various processing vessels without allowing the development of flow instabilities that are typically associated with the "demixing", i.e., the segregation of the cellulosic biomass from the aqueous phase. Here, it is demonstrated that the use of a gelation agent, hydroxypropyl guar gum (HPG) at the relatively low concentration of 0.5 wt% significantly affects the development of viscoelastic material functions of cellulosic suspensions, and improves the dispersive mixing of the cellulose fibers within the aqueous phase. This results in the reduction of the flow instabilities and associated demixing effects that are ubiquitously observed during the pressurization of cellulosic suspensions in Poiseuille and compressive squeeze flows.

1. Introduction

Lignocellulosic biomass has played a key role in daily life for centuries for heating, cooking, clothing and paper manufacture. Lignocellulosic biomass can also serve as an alternate clean energy source upon being converted into ethanol or $H_2$ to reduce reliance on fossil fuels [1, 2]. Sustainable cellulosic biomass encompasses various perennial grasses including those that use C4 biosynthesis, i.e.,سugarcane stover, elephant grass, switch grass and agave. Other sources include woody biomass from lumber industries and excess wood residues from forests [3, 4]. The annual quantity of processed cellulose has reached over $7.5 \times 10^{10}$ tons globally [5].

Lignocellulosic biomass consists of three basic components, i.e., 15-20 wt% of lignin (a non-fermentable phenyl-propene unit), 40-50 wt% of cellulose (a glucose polymer) and 25-35 wt% of hemicellulose (a sugar heteropolymer) plus smaller concentrations of minerals, oils and soluble sugars [6, 7]. The conversion of the lignocellulosic biomass to biofuels usually involves a pretreatment process which includes the removal of lignin or the hemicelluloses, or both, to convert the cellulose so that it is more accessible to treatment. For example, the removal of hemicellulose can be accomplished by hydrolysis in water using conventional vertical or horizontal digesters used in pulp and paper. Sometimes an acid is used to further hydrolyze the cellulose to glucose. Horizontal digesters can also be used to carry out the hydrolysis pretreatment of the biomass and/or steam explosion simultaneously. In steam explosion the biomass is treated with high pressure steam, then it is rapidly depressurized; the depressurization destroys the fibrous structure producing small micron size particles, making the cellulose more accessible to downstream processing such as enzymatic or hydrothermal treatments.

In some processes steam explosion can be carried out downstream of vertical digesters for size reduction. The removal of lignin can effectively be realized via alkaline, NaOH, or ammonia treatment in digesters. It can also be done using reactive extrusion [4, 8]. The pretreatment targeting lignin or hemicellulose removal, is usually followed by a hydrothermal or enzymatic conversion of the cellulose to sugars and fermentation of the sugars to ethanol [1, 9, 10]. Although it is possible to directly convert biomass to ethanol without the conventional pretreatment processes [11, 12] such conversions are currently limited because of the expensive nature of the catalysts and the bacteria used [13]. Regardless of the path of conversion and the type of processing chosen, the cellulosic biomass needs to be mixed with water or solvents to generate suspensions which have to be further pumped and processed in between various treatment vessels [14, 15].

Such pumping and processing can be challenging because of the need to improve processing efficiencies and to reduce costs, while keeping the concentration of the biomass in the suspension relatively high, i.e., in the wt% range of 10 to 20% (volume fraction, $\phi$, range of 0.07 to 0.15) [14, 15]. The processing of suspensions of cellulosic fibers with relatively high aspect ratios (fiber length over the fiber diameter) at such high concentrations is difficult. The fibers can form particle to particle networks which can span the volume of the suspension to induce gel-like behavior, that in turn prevents the break-up of particle agglomerates during mixing and thus limit the dispersion of the cellulose particles within the aqueous matrix.

Furthermore, during subsequent processing the cellulose fibers can readily "demix", i.e., segregate from the aqueous phase either due to sedimentation, given sufficient time, or when the cellulosic suspension is forced to undergo pressure-driven flows, especially through converging geometries (involving the reduction of cross-sectional area available for flow). Such pressure-driven converging flows readily give rise to the development of flow instabilities that manifest themselves by the unsteady nature of the pressures necessary to maintain flow as a consequence of the formation of mats of solids at converging geometries. The formation of mats of solids leads to the filtering of the liquid phase of the suspension and gives rise to the development of concentration gradients of particles in the bulk flow direction [4, 16-18].

In this investigation the shear viscosity and the linear viscoelastic material functions, as well as the mixing and demixing behavior, of three aqueous suspensions of cellulosic fibers (obtained from a commercial bio-refinery) and a model suspension of microcrystalline cellulose (MCC), dispersed into deionized water, were characterized using parallel plate, capillary and squeeze flow rheometers. The effects of the incorporation of a polysaccharide based gelation agent, i.e., a hydroxypropyl guar gum on the mixing and demixing of the cellulose fibers, the associated rheological behavior and the development of flow instabilities and concentration gradients were investigated.

2. Experimental

2.1 Materials

Three lignin-free aqueous suspensions containing 21, 20, and 25 wt% (solid volume fraction, $\phi$=0.159, 0.151 and 0.189) of cellulosic fibers were received from the biomass conversion facility of Renmatix Inc.. It was observed that the initially received shipments of the samples in five gallon containers and the time elapsed between the shipment and receiving (typically one week) gave rise to significant sedimentation effects. Sedimentation was revealed through the formation of a binder-rich layer at the top and the formation of a particle-rich layer at the bottom of the containers. How would one remix such demixed suspensions without altering the compositions of the suspensions? The path that was followed required the reshipment of the lignin-free suspensions using smaller, i.e., 1.5 L containers. This smaller size was conducive to remixing the entirety of the volume of the sample using a Ross planetary mixer with a 2 L mixing bowl. This procedure enabled the conservation of the formulation while allowing remixing.

The remixing of these three suspensions (designated as #1-3) was carried out at a rotational speed of 120 rpm and under ambient temperature in the Ross double planetary mixer. The degree of fill (volume occupied by the suspension over the volume of the mixing bowl) of the Ross mixer was kept at 75%, to allow the effective remixing of the entire as-received suspension sample. A fourth suspension, "a model cellulose suspension," was also prepared by mixing microcrystalline cellulose (MCC) purchased from Sigma-Aldrich (Catalog#: 435236, Mo, USA) with DI water at a particle concentration of 30 wt% (solid volume fraction, $\phi$=0.227).

It is very important to document the efficiencies of the mixing and remixing processes and the resulting distributions of the particle concentrations within the aqueous matrix to allow the generation of day-to-day reproducible mixtures with similar levels of homogeneity of the concentrations of the ingredients of the formulation [19-23]. It is generally difficult to mathematically model the dynamics of the mixing process to identify robust mixing geometries and conditions [24-29] and experimental means are preferred for the characterization of measures of "goodness of mixing" to guide the mixing process so that reproducible mixtures can be generated, as described next.

### 2.2 Mixing index characterization

The following analysis was used to characterize the statistics of the concentration distributions of the cellulose particles to provide measures of the "degree/goodness of mixedness" or "mixing indices" [20, 23]. If N measurements of the concentration, $c_i$, of one of the ingredients of the suspension formulation are made, then the mean, $\bar{c}$, and the variance, $s^2$, of the concentration distribution of this particular ingredient can be obtained, i.e., $\bar{c} = \frac{1}{N}\sum_{i=1}^{N} c_i$ and $s^2 = \frac{1}{(N-1)}\sum_{i=1}^{N}(c_i - \bar{c})^2$. A small variance value would suggest that the mixture approaches the behavior of a homogeneous system, where most of the samples yield concentration values, $c_i$, that approach the mean concentration, $\bar{c}$. On the other hand, the poorest mixing state would pertain to the components of a mixture being completely segregated from each other. The value of the maximum variance $s_0^2$ for a segregated system can be defined by assuming that the samples are taken from either one component or the other without crossing a boundary [19, 30] and the degree of mixedness, i.e., mixing index, *MI*, can be accordingly defined using the ratio of the standard deviation of the concentration distribution over the standard deviation of the completely segregated sample as:

$$s_0^2 = \bar{c}(1-\bar{c}) \quad \text{and} \quad MI = 1 - s/s_0 \qquad \text{Eq. (1)}$$

Thus, the value of the mixing index, *MI*, would approach one for a completely random distribution of the ingredients and zero for the segregated state.

For the mixing index determination the suspension samples were subjected to thermo gravimetric analysis, TGA, using a TA Instruments Q50. Five specimens of around 20 mg each were collected from different locations inside the mixer. Each specimen was heated at a rate of 20 °C/min from ambient to 110 °C and held at this temperature for 30 min. Weight changes of the specimens were recorded as a function of time and temperature. Consequently, the total concentration variation of the cellulose fibers was used for the determination of the mixing indices.

For Suspensions #1-3 that were remixed in the Ross mixer a mixing index > 0.98 could be reached after 30 minutes of remixing. Attainment of this mixing index was considered to represent a well-mixed state. For the mixing of the MCC into DI water (Suspension #4) the mixing was carried out using an overhead propeller mixer. The mixing volume was 200 mL and the degree of fill of the mixer was kept at 50%. For the mixing of MCC into DI water 20 minutes of mixing was sufficient to reach a mixing index of 0.98.

**Gelation agent**

All four suspensions were also incorporated with a polysaccharide, hydroxypropyl guar gum (HPG). HPG is a well known nonionic biopolymer used as a gelling agent. It was procured from Solvay Inc. with the trade name of Jaguar HP 8 COS (CAS number 39421-75-5). Due to its intermolecular hydrogen bonding ability, HPG can aggregate into ribbon-like structures without a crosslinker present [31]. This leads to hydrogels with relatively high viscosities even at modest concentrations [32]. The gelation agent was applied at three concentrations of 0.1, 0.5 and 1 wt%. A sonicator horn was used to incorporate the gelation agent into the suspension samples. A sonication duration of 10 minutes was determined to be adequate on the basis of repetitive measurements of dynamic properties (frequency-dependent linear viscoelastic moduli obtained upon small-amplitude oscillatory shear and this sonication time was kept constant for all the specimens.

**2.3    Analysis of cellulose fiber lengths and diameters**

For the determination of the sizes of the as-received cellulose fiber aggregates (there was significant fiber agglomeration), and the MCC fiber aggregates the remixed suspensions were diluted to 1 wt% and then dried using a vacuum oven at 110 °C overnight. Upon drying, the micrographs of the fibers were obtained using a Zeiss Auriga Dual-Beam FIB-SEM (Carl Zeiss Microscopy) at an accelerating voltage of 1kV. Typical micrographs of the aggregates for Suspensions #1-4 are shown in the insets of Figure 1.

The cellulose fiber aggregates obtained prior to mixing in the Ross mixers were characterized for their diameter and length distributions using ImageJ software. At least 10 micrographs were analyzed for each suspension. The typical diameters of fiber aggregates of Suspensions #1-3 were in the 12-15 μm range, whereas the diameters of the MCC aggregates of Suspension #4 were around 35 μm. The cumulative length distributions of the fiber aggregates are shown in Figure 1. The mean values of the fiber dimensions are also provided in Table 1. Suspension #1 exhibited fiber aggregate lengths that were in the 20 to 200 μm range, with 80% smaller than 100 μm. Suspension #2 exhibited aggregate lengths in the 20 to 300 μm range with 80% smaller than 150 μm. Suspension #3 exhibited the broadest fiber length distribution, i.e., the fiber lengths ranged from 40 to 400 μm, with 80% smaller than 200 μm.

Finally, the cellulose fibers of Suspension #4 exhibited the narrowest length distribution, i.e., with 90% of the fiber lengths falling in the range of 70 to 170 µm. The mean aspect ratios (ratios of the length over the diameter) of the aggregates of Suspensions #1-4 were around 3-7. It can be assumed that these relatively low aspect ratio aggregates will undergo exfoliation during the mixing or remixing processes to generate fibers with greater aspect ratios.

### 2.4 Rheological characterization

The dynamic properties, steady torsional flow and the capillary flow behavior of the four fibrous suspensions were characterized using an Advanced Rheometric Expansion System (ARES) rotational rheometer and an Instron capillary rheometer, respectively. The rotational rheometer was equipped with a force rebalance transducer 0.2K-FRTN1. The rheometer was used with stainless steel parallel disks with 12.5-25 mm radii, $R$. The torque accuracy of the transducer was ±0.02 g-cm. The actuator of the ARES is a DC servo motor with a shaft supported by an air bearing with an angular displacement, $\theta$, range of $5\times10^{-6}$ to 0.5 rad, and angular frequency, $\omega$, range of $1\times10^{-5}$ to 100 rad/s. The angular rotational speed, $\Omega$, can be varied between $1\times10^{-6}$ to 200 rad/s. Parallel disk geometry was employed to carry out the tests at ambient temperature. The gap, $H$, between the two parallel disks was kept constant at 1 mm. The sample loading procedure was kept the same for all the experiments (loading a constant volume of the sample via the use of a shim cavity). Consistent with our earlier investigations the suspension samples were not presheared. Preshearing itself can significantly change the structure and hence the resulting rheological behavior of concentrated suspensions [33].

The schematic diagram of rotational rheometry is shown in Figure 2a. During oscillatory shearing the shear strain, $\gamma$, at the edge varies sinusoidally with time, $t$, at a frequency of $\omega$, i.e., $\gamma(t) = \gamma^0 \sin(\omega t)$ where $\gamma^0 = \frac{\theta R}{H}$ is the strain amplitude at the edge ($\theta$ is the angular displacement). The shear stress at the edge, $\tau(t)$, response of the fluid to the imposed oscillatory deformation consists of two contributions associated with the energy stored as elastic energy and energy dissipated as heat, i.e., $\tau(t) = G'(\omega)\gamma^0 \sin(\omega t) + G''(\omega)\gamma^0 \cos(\omega t)$. The storage modulus, $G'(\omega)$ and the loss modulus, $G''(\omega)$, also define the magnitude of complex viscosity, $|\eta^*| = \sqrt{(G'/\omega)^2 + (G''/\omega)^2}$, and $\tan\delta = G''/G'$. In the linear viscoelastic region all dynamic properties are independent of the strain amplitude, $\gamma^0$.

Special attention was paid to reduce the water evaporation rate during the small-amplitude oscillatory shearing experiments. This was achieved by placing a solvent trap around the sample. Time sweeps of 10 minutes at constant frequency and strain amplitude indicated that the suspensions were stable throughout the characterization of their linear viscoelastic material functions as a function of $\omega$.

The schematic diagram of capillary rheometry is shown in Figure 2b. An Instron Floor Tester was used in conjunction with a temperature-controlled reservoir (the "barrel"). An extrusion ram (the "plunger"), pressurized the suspension sample held within the reservoir and ram-extrude it through a capillary installed at the bottom of the reservoir. The plunger was connected to a cross-head bar traveling at constant velocity and housed a normal force transducer "the load cell". The load cell allowed the monitoring of the time-dependent development of pressure drop at various cross-head speeds (hence at various volumetric flow rates). The capillary flow experiments employed a capillary with a radius, $R$, of 1.25 mm and a length, $L$, of 100 mm, generating a length/radius ratio of 80. This $L/R$ ratio was considered high enough to assume that entrance and exit losses were negligible in comparison to the pressure drop associated with fully-developed flow in the capillary. The barrel of the rheometer had a radius of 4.75 mm and thus the contraction ratio (diameter of the barrel over the diameter of the capillary die) was 3.8. It is the converging flow associated with this contraction ratio that can test the stability of the flowability of concentrated suspensions.

During capillary flow experiments the barrel was filled with 5-10 g of suspension and the cross-head was run at a constant speed of 0.53 mm/s while documenting the time-dependent development of the normal force. The capillary flow experiments specifically investigated the development of flow instabilities associated with the segregation of the binder and the particles from each other. This happens as a consequence of the formation of particle mats at the converging flow region, to force the axial migration and filtration of the binder. Such segregation and binder filtration effects could be documented earlier for converging flows of concentrated suspensions [16, 17].

An additional rheometer, i.e., the squeeze flow rheometer was also used to test the flowability of the suspensions. Figure 2c shows the schematic diagram of the squeeze flow rheometer. A load cell which measures the time-dependent development of the normal force was connected to a plunger. The plunger was connected to a cross-head, the velocity of which could be controlled via a stepper motor. Squeeze flow rheometry is a very convenient viscometric flow for concentrated suspensions [18, 34, 35]. During squeeze flow suspension specimens of 3.5 g were placed within two mating splittable sample holders to generate disk-shaped specimens with a diameter of 25 mm. The ram was immediately brought down and squeezed the disk-shaped sample at a constant speed of 0.09 mm/s and the resulting normal force versus time data were collected.

### 2.5 Characterization of particle concentration distributions

For both capillary and squeeze flow experiments 200 mg specimens were collected as a function of location and time. For capillary rheometry this amounted to the collection of extrudate samples emerging out of the capillary rheometer as a function of time, as well as the collection of location-dependent samples from the barrel of the rheometer following 2 minutes of extrusion. For squeeze flow the sample collection procedure involved the dead stop of the squeeze flow followed by the collection of specimens from various radial locations of the squeezed sample. All samples were weighed immediately upon collection and were placed in a vacuum oven at 110 °C for drying. During drying the sample weights were documented until no further weight changes could be detected and the weight differences before and after drying were used for the reporting of the concentrations of cellulose fibers.

### 3. Results and discussion

Figures 3-6 show the storage modulus, $G'$, loss modulus, $G''$, and the magnitude of the complex viscosity, $|\eta^*|$, values as functions of frequency, $\omega$, of Suspensions #1-4 with and without the incorporation of the gelation agent, HPG (HPG concentrations of 0.1, 0.5 and 1 wt%). For all suspensions the strain amplitudes, at which linear viscoelastic behavior prevailed,

were restricted to relatively small strain amplitudes, $\gamma^0$, i.e., $\gamma^0 \leq 0.1\%$. Without the incorporation of HPG, the $G'$ and $G''$ values of suspensions #1-4 exhibited little dependence on frequency. The $G'$ values were about one order of magnitude greater than the $G''$ values. These two observations suggest that the four suspensions of cellulose fibers of our investigation all exhibit gel-like behavior [36, 37]. Such gels are typically viscoplastic and exhibit yield stresses [18, 23, 38]. The observed gel-like behavior in the absence of a gelation agent occurs due to the formation of a particle to particle network.

Similar gel-like behavior is observed with suspension samples upon incorporation of 0.1 wt% of HPG (Figures 3-6). The gelation agent at 0.1 wt% increases both the energy stored as elastic energy and the energy dissipated as viscous energy dissipation during one cycle of deformation as represented by the increased values of $G'$ and $G''$ upon the incorporation of the gelation agent. For example, the $G'$ and $G''$ values of Suspension #1 increased by about 50% when the gelation agent was incorporated at 0.1 wt%. The independence of the dynamic properties from frequency and $G' \gg G''$ again indicate that a relatively strong fiber to fiber network had been established. Furthermore, the incorporation of the gelation agent at 0.1 wt% increased the gel strength i.e., the plateau values of $G'$ and $G''$ increased. The increase of the gel strength of a gel-like suspension following the incorporation of a gelation agent is the expected behavior. Typically the $G'$ and $G''$ values of the suspensions would increase with the increasing $G'$ and $G''$ values of the liquid phase of the suspension [31, 32].

However, when the gelation agent HPG was incorporated at 0.5 and 1 wt% the $G'$ and $G''$ values of the resulting suspensions decreased significantly and became more sensitive to frequency (Figures 3-6). The decreases of the dynamic properties upon the incorporation of the gelation agent into the formulation at 0.5 and 1 wt% became more prominent at the lower end of the frequency scale at which the dynamic properties would be more sensitive to the structure of the suspensions. What has given rise to the decrease of the dynamic properties, i.e., to the decrease of the viscosity and elasticity upon the incorporation of the gelation agent at the higher concentrations of 0.5 and 1 wt%?

To answer this question, the role the gelation agent plays on the dynamic properties of the aqueous phase was investigated first. The effects of the concentration of the gelation agent on the dynamic properties of the binder (DI water) are shown in Figures 7 and 8. The dynamic properties increase significantly with increasing gelation agent concentration to indicate that the elasticity and the viscosity of the liquid phase ("the binder") increase with the increasing concentration of the gelation agent. As would be expected the dynamic properties, i.e., $G'$ and $G''$ and the magnitude of complex viscosity, $|\eta^*|$, increase monotonically with the increasing concentration of the gelation agent. Generally, with the increase of the elasticity and the viscosity of the binder phase, there would be a corresponding increase of the elasticity and the viscosity of the suspension. If this is so, why would the elasticity and viscosity of the suspensions decrease with the incorporation of the gelation agent at 0.5 and 1 wt%?

It is hypothesized here that the observed decreases in elasticity and viscosity of the cellulose suspensions with increasing concentrations of the gelation agent are related to the enhanced efficiency of the dispersive mixing process due to the presence of the gelation agent at the higher concentrations. Specifically, the increase of the dispersive mixing efficiency due to the increase of the elasticity and viscosity of the binder phase could lead to decreases in the sizes of particle agglomerates thus giving rise to decreases in the elasticity and the viscosity of the suspensions. The decreases in particle agglomerate sizes can be related to the increases in the stress magnitudes that are applied on the suspensions with increasing binder viscosity and elasticity (Figures 7 and 8). When the applied stress magnitudes overcome the cohesive strengths of the fibers of the cellulose aggregates (that are typically held together with relatively weak forces) the aggregates rupture [19].

The cellulose agglomerates consist of aggregated cellulose fibers with extensive pores capable of holding relatively large amounts of water by capillarity [39, 40]. As the specific energy input (mechanical energy over time and volume) during the mixing or remixing processes increases with the time of mixing the agglomerates would break down releasing the relatively significant quantities of water that are held between the fibers of the aggregates. This should give rise to an increase in the volume fraction of the binder (and thus to the decrease of the volume fraction of solids) and consequently result in the decreases of the elasticity and shear viscosity of the suspensions, as manifested here via the decreases of the storage and loss moduli, and their magnitude of complex viscosity values when the concentration of the gelation agent, HPG, was increased to 0.5 and 1 wt% (Figures 7 and 8).

Similar decreases in dynamic properties and viscosity with increasing specific energy input during mixing has also been observed with various concentrated suspensions with polymeric binders [20, 22]. However, it is interesting to note that for cellulose suspensions there appears to be a critical concentration of the gelation agent (0.5 wt%) that is necessary for enabling the breakdown of the agglomerate structure to occur during the mixing process.

**Effects of the gelation agent on the stability of the ram extrusion and squeeze flow processes**

Figures 9a to 9d show the time dependencies of the wall shear stress, $\tau_w = \frac{\Delta P}{2L}R$, for Suspensions #1-4, respectively. Here $\Delta P$ is the total pressure drop at the apparent shear rate at the wall, $\dot{\gamma}_a = \frac{4Q}{\pi R^3}$, $Q$ is the volumetric flow rate at the constant ram speed of 0.53 mm/s for the capillary die with a 1.25 mm radius, $R$, (apparent shear rate=24.6 s$^{-1}$) and length of 0.1 m.

As shown in Figure 9a the wall shear stress increased monotonically during the capillary flow of Suspension #1 without the gelation agent at an apparent shear rate of 24.6 s$^{-1}$. The wall shear stress reached $8\times10^5$ Pa in 60 seconds. The monotonic increase is indicative of the unsteady state nature of the capillary flow process that occurred for Suspension #1. Similar unbounded increases of the wall shear stress at the same apparent shear rate of 24.6 s$^{-1}$ were observed for Suspensions #2-4 without the gelation agent present, as shown in Figures 9b, 9c and 9d. The wall shear stress increased rapidly and reached values as high as $7\times10^5$, $3\times10^5$, and $9.5\times10^5$ Pa for Suspensions #2-4, respectively. These unbounded increases in pressure are indicative of flow instabilities associated with the demixing processes that are taking place. It is considered that the demixing occurs as a consequence of the converging flow taking place in capillary rheometry as explained next. The primary culprit in the development of such unbounded increases in pressure drop during extrusion of concentrated suspensions in capillary flow is the formation of a mat of solids at the converging section of the capillary rheometer [16-18, 41, 42]. The mat formation involves the interlocking of particles at the converging region, and the resulting blockage of the bulk flow. Since the plunger moves at a constant rate during ram extrusion in capillary rheometry the mat of solids formed at the converging section of the capillary rheometer serves as a filter bed leading to the axial migration of the binder phase (giving rise to further increases of the pressure drop with diminishing binder concentration in the reservoir).

The effects of the incorporation of the gelation agent on the stability of the capillary flow were striking. For all four suspensions steady flow was achieved and could be maintained for the gelation agent concentrations of 0.5 wt% and above (results for 1 wt% are not shown). On the other hand, steady flows could not be achieved for any of the four suspensions when the gelation agent was missing or when it was incorporated at the lower concentration of 0.1 wt%.

The mechanisms of the formation of a mat of cellulose fibers and the resulting axial migration of the binder phase during capillary flow were further elucidated via the measurements of the particle concentrations in the extrudates emerging from the capillary flow (Figure 10a shows the schematic diagram of the capillary flow.) The extrudates were collected as a function of time during extrusion and the cellulose concentrations of the extrudate sections were determined. Furthermore, the cellulose particle concentration distribution as a function of location in the barrel was also determined following the dead stop of the ram extrusion process. The segregation of the binder could be clearly demonstrated via the images of the extrudates collected during the extrusion process (Figure 10b). The first image in Figure 10b was collected for a suspension sample containing 0.1 wt% of the gelation agent, HPG within 10 s of the initiation of the extrusion process. The image shows that the extrudate is transparent, i.e., it consists of the binder only, indicating that the liquid phase was filtered out. On the other hand, extrudates with no visible separation of the binder were observed when the gelation agent was increased to 0.5 wt% (as demonstrated with the second image in Figure 10b).

The particle concentrations remaining in the barrel as a function of distance from the entrance to the capillary die are shown in Figure 10c (upon a dead stop following an extrusion time of 170 s) for different concentrations of the gelation agent, HPG. The cellulose fiber concentration reached values as high as 45 wt% at locations close to the capillary entrance (1-9 mm away from the capillary entrance) after an extrusion time of 170s for Suspension #4 without the gelation agent or with only 0.1 wt% of the gelation agent. This cellulose concentration is significantly higher than the as-mixed suspension, i.e., 30 wt% of cellulose. It is apparent that the binder content of the suspension has been diminished upon its filtration based axial migration, resulting in the increase of the cellulose concentration in the barrel during the elapsed extrusion time. It is also interesting to note that without the gelation agent present and the minimum gelation agent concentration of 0.1 wt%, the specimens collected 12 mm away from the die entrance exhibited a solids concentration of only 20 wt%. Thus, a significant binder concentration variation occurred within the barrel (Fig. 10c) in conjunction with the monotonic increase of the extrusion pressure and the changing nature of the extrudates emerging from the capillary (Fig. 10b).

Figure 10d shows the cellulose fiber concentration of the extrudates emerging from the capillary as a function of extrusion time. Without the gelation agent present or at the gelation agent concentration of 0.1 wt% the cellulose fiber concentration is only about 10 wt% at an extrusion time of about 45 s. The cellulose concentration increases monotonically with extrusion time to reach about 45 wt% after 170 s. On the other hand, when the gelation agent is incorporated at a concentration of 0.5 wt% by weight the cellulose fiber concentration remains constant at 30 wt%. The constancy of the cellulose concentration at 30 wt% (Figure 10d) points to the stable nature of the extrusion process for Suspensions #1-4 at the gelation agent concentration of 0.5 wt%.

A number of different mechanisms could give rise to the segregation of the binder liquid from the fibrous particles of the suspension including the shear-induced self-diffusion of curved or straight fibers [43, 44]. As noted earlier sedimentation of the fibrous phase can also occur. Sedimentation effects generally become significant if the density difference between the particles and the binder, $\rho_s - \rho_b$, is relatively high. The sedimentation velocity of the particles incorporated into a Newtonian fluid with viscosity, $\mu_b$, can be given as: [45]

$$u_{sedimentation} = \frac{D_p^2 g (\rho_s - \rho_b)}{18 \mu_b} \qquad \text{Eq. (2)}$$

where $D_p$ is the particle diameter and $g$ is the gravitational acceleration. Although sedimentation is likely to be a significant effect for Newtonian binders when the density difference between the solid particles and the binder fluid, i.e., $(\rho_s - \rho_b)$, is appreciable, the sedimentation effect significantly diminishes for gel-like suspensions with a yield stress, $\tau_y$. The critical ratio of yield stress over the driving buoyancy stress, i.e., the yield parameter, $Y$, is defined as $Y = \frac{\tau_y}{g D_p (\rho_s - \rho_b)}$. Various studies have suggested that there would be no motion of the spherical particles when the yield parameter, $Y$, is > 0.02-0.05 [46]. Thus, gel-like behavior clearly reduces the tendency to segregate and the sedimentation effect would not be a relevant mechanism for segregation of the binder from the particles within the relevant time scale of our capillary flow experiments.

A second segregation mechanism is associated with the formation of a mat of solids at converging flow regions with the mat acting as a filter to drive the segregation, i.e., the filtration, of the binder from the particle mat [16, 17]. During steady capillary flow conditions the pressure drop imposed on the suspension $\Delta P$, generates a flow with a wall shear stress of $\tau_W = \frac{\Delta PR}{2L}$, associated with a bulk axial velocity of $V_b$. Furthermore, the pressure drop $\Delta P$ can also generate a filtration velocity, $\bar{V}_m$, i.e., $\bar{V}_m = \frac{\bar{V}_0}{\varepsilon}$ where $\bar{V}_0$ is the superficial velocity of the binder liquid within the mat of fibers acting as a filter bed and $\varepsilon$ is the void volume related to the volume fraction of particles, $\phi$, i.e., $\varepsilon = 1-\phi$. The filtration velocity of the binder in a cylindrical flow channel can be analyzed via the mechanism of filtration through packed beds [47]. For steady laminar flow of a Newtonian liquid the average filtration velocity $\bar{V}_m$ (based on the tube diameter) is given as follows:

$$\bar{V}_m = \frac{D_p^2}{150\mu_b L}\left(\frac{1}{\phi}-1\right)^2 \Delta P \qquad \text{Eq. (3)}$$

Previous investigations have indicated that a critical shear stress at the wall, $\tau_{cr}$, exists. At wall shear stresses that are smaller than the critical shear stress at the wall, $\tau_{cr}$, $\bar{V}_m$ is greater than the bulk velocity $V_b$ and consequently filtering and unstable flow would occur [Yilmazer et al., 1989; Yaras et al., 1994]. The critical shear stress at the wall, $\tau_{cr}$, is related to the suspension properties as:

$$\tau_{cr} \propto \frac{D_p^2}{\mu_b R}\left(\frac{1}{\phi}-1\right)^2 \qquad \text{Eq. (4)}$$

Eq. (4) suggests that increases in the matrix viscosity, $\mu$, and the tube radius, $R$, and decreases in the filler particle size, $D_p$, will shift $\tau_{cr}$ to smaller values and thus promote stable flows. Clearly in our experiments the increase of the binder viscosity via the incorporation of the gelation agent has indeed promoted the achievement of a stable flow consistent with Eq. (4).

The squeeze flow experiment was next used to document the effects of the gelation agent on the demixing of cellulose suspensions during pressurization. Figure 11a shows the normal force generated versus the reciprocal gap during squeeze flow at a constant ram speed of 0.09 mm/s for Suspensions #1-3 with and without 0.5 wt% of the gelation agent. All three suspensions exhibited exponential growth of the normal force with time during squeezing. Normal forces were negligible initially but eventually reached the load cell upper limit of 6500 N. Suspensions #1, #2 and #3 reached the load cell normal force limit at gaps of 0.6, 0.5 and 0.45 mm (reciprocal gaps of 1.7, 2.0, 2.2 mm$^{-1}$), respectively. On the other hand, all three suspensions exhibited negligible normal force values in the squeeze flow tests within the gap range of 1.5 to 0.4 mm, when they were incorporated with 0.5 wt% of HPG.

Figure 11b shows the typical cellulose concentration distributions for Suspension #2 along the radial direction achieved upon the completion of the squeeze flow. The concentration distributions are very different for the suspension without the gelation agent versus the suspension with 0.5 wt% gelation agent. Without the gelation agent present, the cellulose concentration reaches 65 wt%, i.e., a cellulose concentration that is almost three times higher than that of the as-mixed suspension, at a location of 5 mm from the axis of symmetry. On the other hand, at a radial distance of 20 mm from the axis of symmetry, the cellulose concentration is only 17 wt%. Thus, significant variations in the concentrations of the binder and the cellulose fibers develop upon the squeezing of the suspension when there is no gelation agent present.

The inset shown in Figure 11b is an image of the as-squeezed sample. The differences in the water content as a function of the radial position are apparent from the image. A dramatic liquid phase migration appears to have occurred along the radial direction leading to the depletion of the binder phase at locations close to the axis of symmetry. The binder-rich nature of the first compressed (and squeezed out of the gap) suspension is also evident and is consistent with the cellulose concentration distribution shown in Figure 11a. The filtration of the binder phase during squeeze flow of concentrated suspensions has been reported by others [48-50] on the basis of mechanisms that were documented for converging flow [16, 17]. In these earlier investigations the binder phase was observed to exhibit a relative motion compared with the bulk of the suspension due to the high permeability of the filter cake formed by the squeezed fibers [50]. It is clear that this mechanism is similar to the mechanism of development of flow instabilities, associated with the formation of a mat of solids at converging flow geometries, and the filtration of the binder through the mat which acts as a filter bed [16, 17].

The effects of the incorporation of the gelation agent at 0.5 wt% on the stability of the squeeze flow are again striking as shown in Fig. 11a. When the gelation agent was incorporated at 0.5 wt% the cellulose concentration at different radial locations were identical. This suggests that the cellulose suspension remained homogeneous during the squeezing flow when the gelation agent was incorporated at ≥ 0.5 wt%. Altogether this finding is consistent with the results of the capillary flow experiments which suggested that the stable flowability of the cellulose suspensions could be achieved when the gelation agent was incorporated at 0.5 wt%.

These results for the converging and squeeze flows of suspensions without the gelation agent and with the minimum gelation agent concentration of 0.1 wt% are harbingers of significant processing issues that would arise during the conversion of the biomass suspensions at the plant scale. These issues were indeed observed during the processing of these suspensions at the industrial scale especially during pumping and mixing. Typically, clogging of flow channels with cellulose fiber mats and pressure spiking were observed and led to difficulties and inconsistencies in process control and the maintenance of product quality. The role of the gelation agent at concentrations ≥ 0.5 wt% is very significant since at these concentrations the dispersive mixing of the cellulose suspensions becomes easier, while the flow instabilities associated with segregation of the binder from the solid phase diminish in importance, enabling stable continuous processability.

4. **Conclusions**

Mixing of cellulosic fibers into aqueous solutions and keeping the cellulosic fibers homogenously suspended are important industrial challenges during the processing of cellulosic biomass. This investigation focusing on the development of viscoelastic material functions and capillary and squeeze flow behavior has revealed that both the dispersion of the cellulose fibers in the aqueous phase and the processability of the cellulosic suspensions can be positively impacted by the incorporation of a gelation agent, provided that the gelation agent, hydroxypropyl guar gum is incorporated above a critical concentration of 0.5 wt%. These findings provide a better understanding of the flow instabilities that occur at the industrial scale and can be used to impart better flowability and processability of cellulosic suspensions during processing.

**Acknowledgements**

We thank Dr. Seda Aktas for her help and input.

**Funding**

This work was supported by Renmatix Corporation, for which we are grateful.

**Figures**

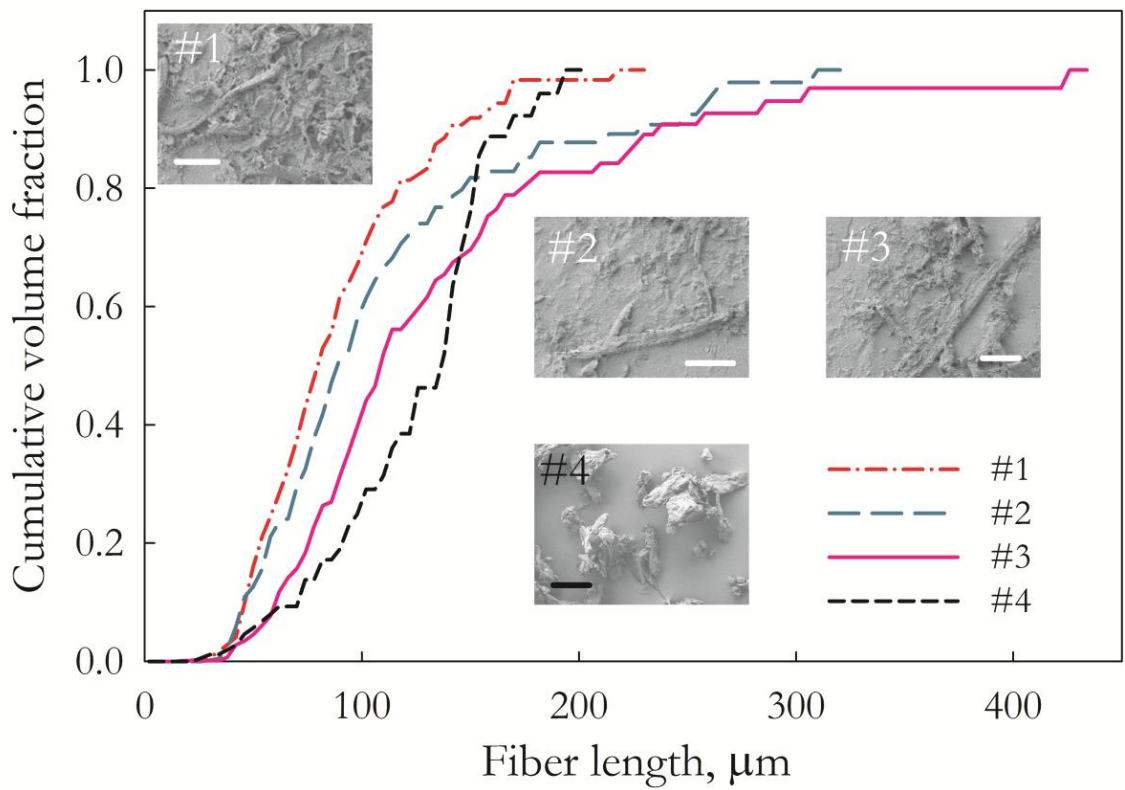

Figure 1. The cumulative volume fraction versus fiber length for four cellulose suspensions. Insets show microscopy images of cellulose agglomerates from each suspension. The scale bar is 40 µm.

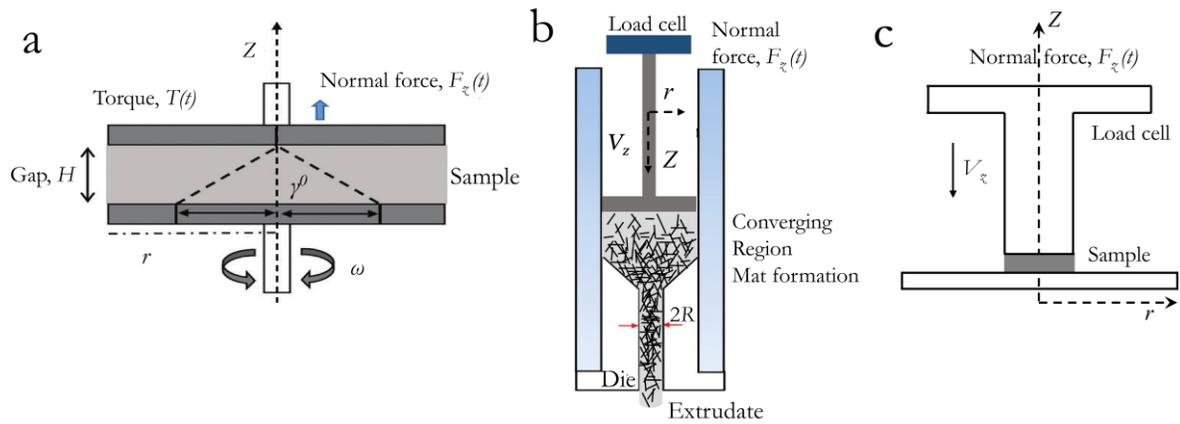

Figure 2. Schematics of the three viscometric flows used for the rheological characterization of the cellulose suspensions a) Small-amplitude oscillatory shear, b) Capillary flow, c) Squeeze flow.

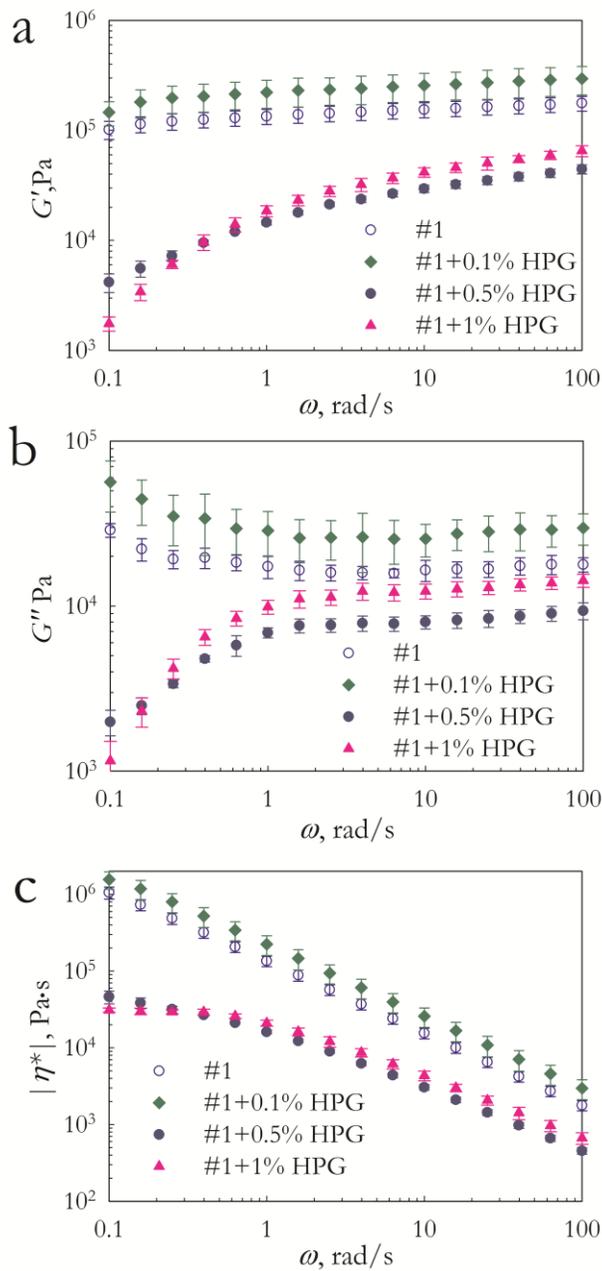

Figure 3. a) Storage modulus, $G'$ b) loss modulus, $G''$, c) magnitude of complex viscosity, $\left|\eta^*\right|$ versus frequency, $\omega$, of Suspension #1 with various concentrations of the gelation agent, HPG. The strain amplitude is 0.1% and the temperature is ambient.

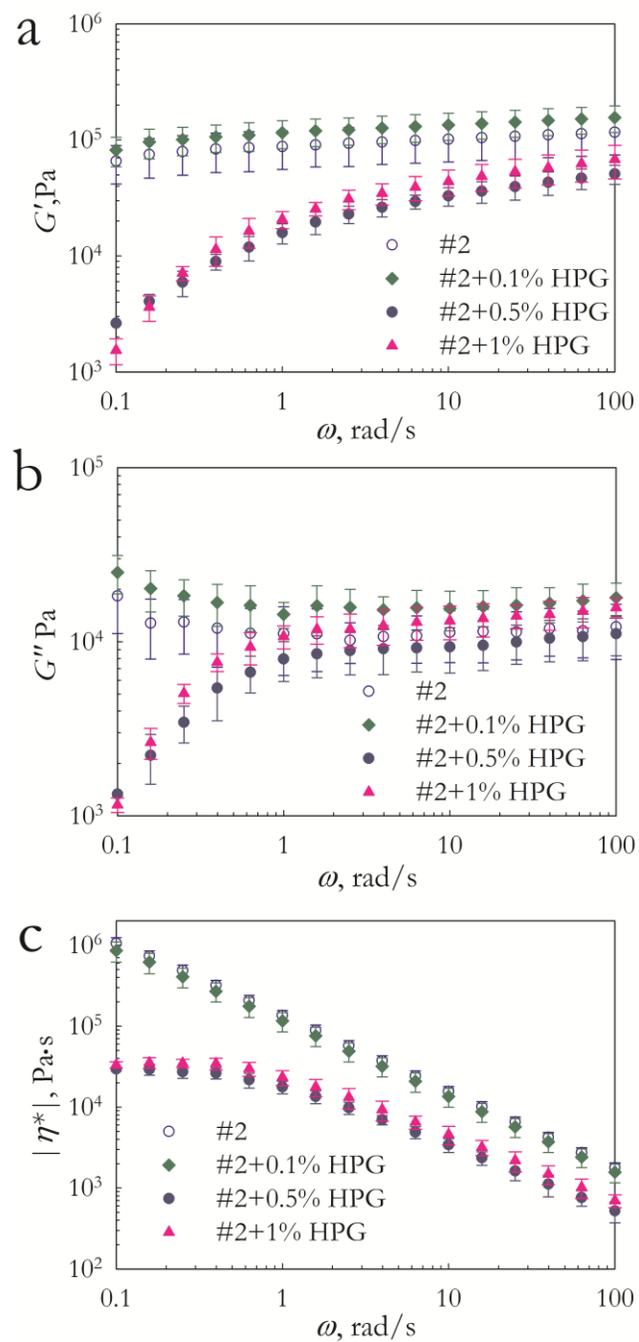

Figure 4. a) Storage modulus, $G'$ b) loss modulus, $G''$, c) magnitude of complex viscosity, $|\eta^*|$ versus frequency, $\omega$ of Suspension #2 with various concentrations of the gelation agent, HPG. The strain amplitude is 0.1% and the temperature is ambient.

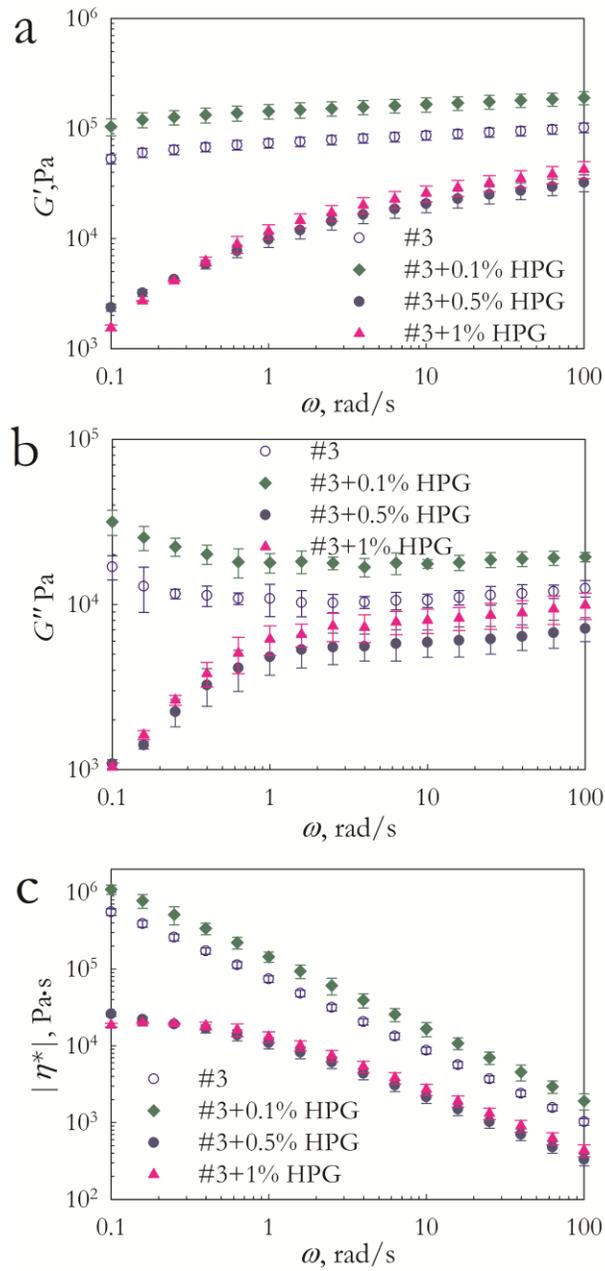

Figure 5. a) Storage modulus, $G'$ b) loss modulus, $G''$, c) magnitude of complex viscosity, $|\eta^*|$ versus frequency, $\omega$ of Suspension #3 with various concentrations of the gelation agent, HPG. The strain amplitude is 0.1% and the temperature is ambient.

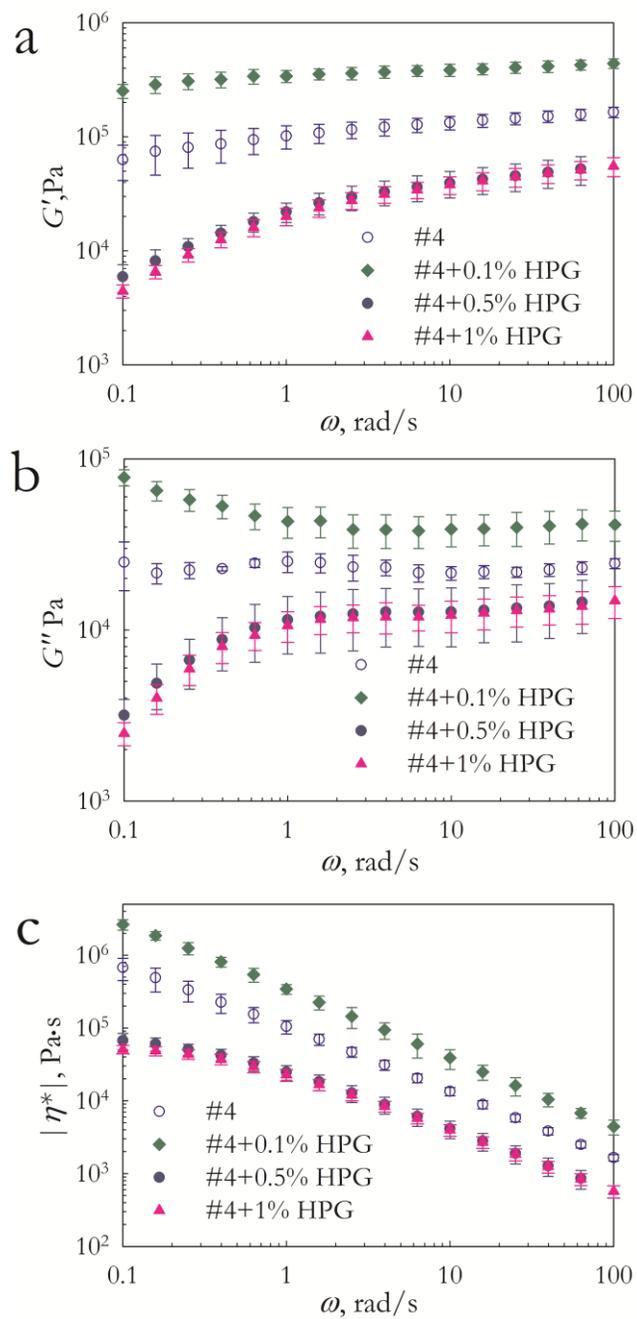

Figure 6. a) Storage modulus, $G'$ b) loss modulus, $G''$, c) magnitude of complex viscosity, $|\eta^*|$ versus frequency, $\omega$ of Suspension #4 with various concentrations of the gelation agent, HPG. The strain amplitude is 0.1% and the temperature is ambient.

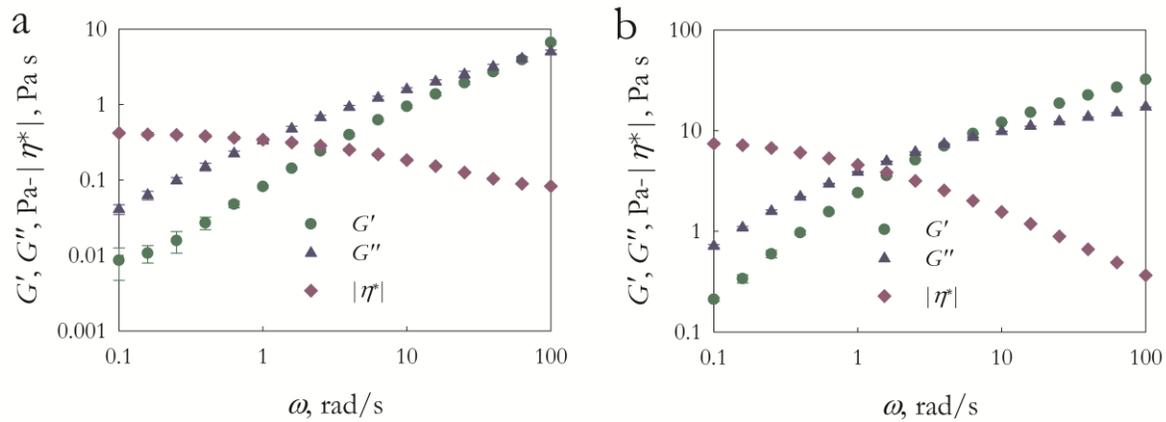

Figure 7. Dynamic properties of DI water with a) 0.5 wt%; b) 1 wt% of the gelation agent, HPG.

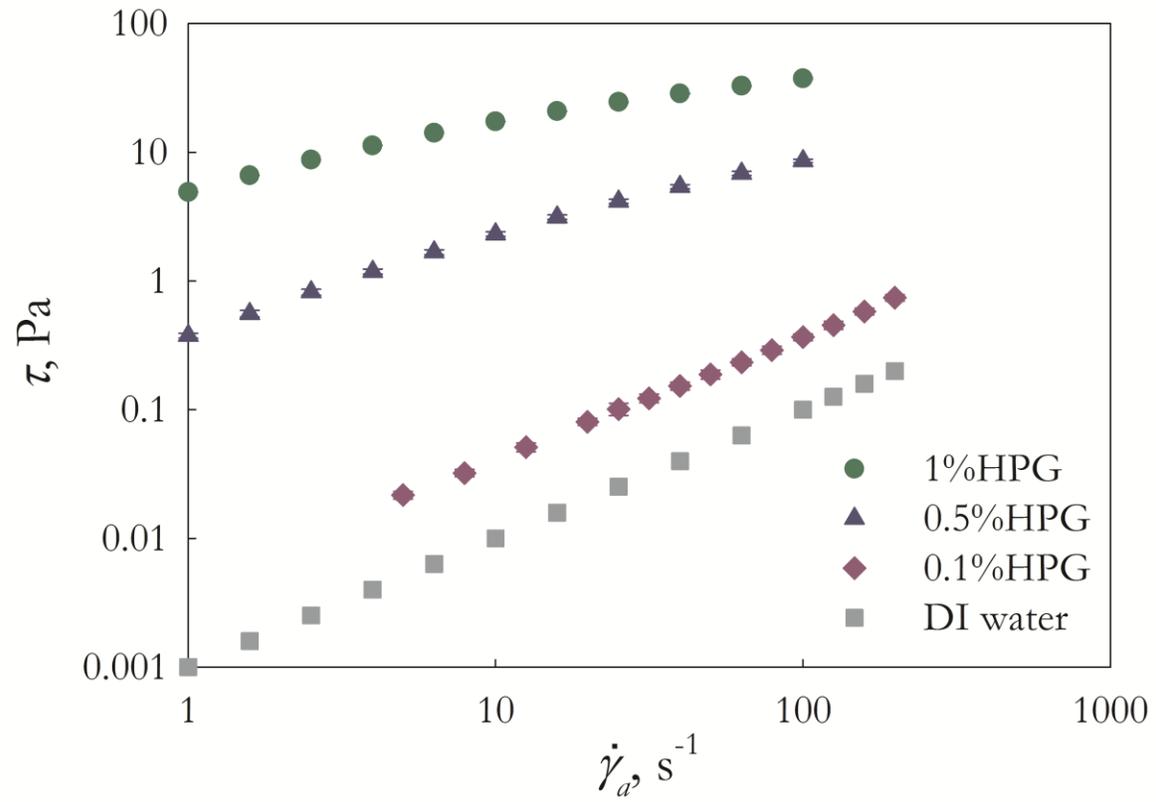

Figure 8. The flow curves (shear stress versus the shear rate) for DI water with 0.1, 0.5 and 1 wt% of the gelation agent, HPG (the flow curve for water with Newtonian viscosity of $10^{-3}$ Pa-s included for reference).

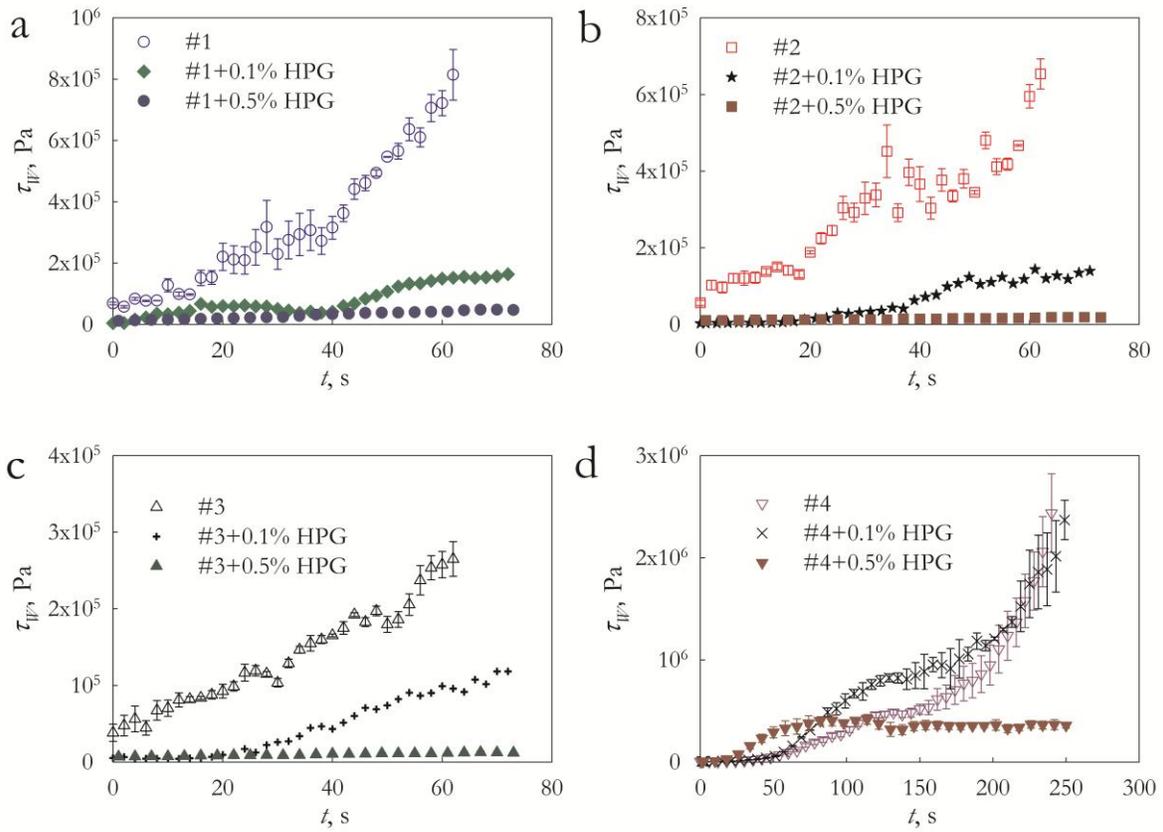

Figure 9. Wall shear stress versus time obtained from capillary flow upon extrusion with a capillary die with a diameter of 2.5 mm and a length/diameter ratio of 40 at an apparent shear rate of 24.6 s$^{-1}$ for various concentrations of the gelation agent, HPG for a) Suspension #1, b) Suspension #2, c) Suspension #3 and d) Suspension #4.

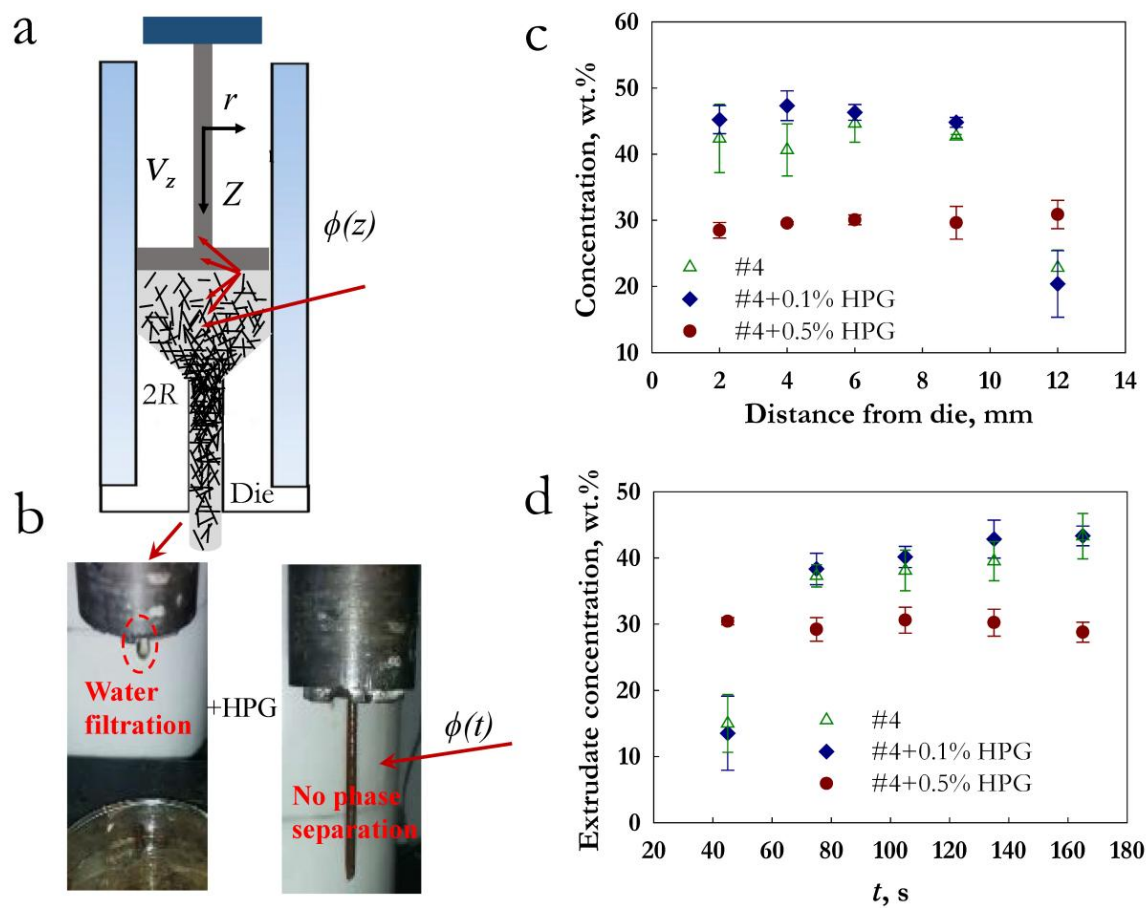

Figure 10. a) Schematics of capillary flow, b) Images of the extrudates of Suspension #2 with and without 0.5 wt% of HPG, c) Cellulose concentration of the Suspension #4 incorporated with various concentrations of HPG collected from various locations in the barrel of the capillary rheometer upon reaching a dead stop at 170 s from the onset of extrusion. d) Cellulose concentration versus time for Suspension #4 of the extrudates collected as a function of time during capillary flow.

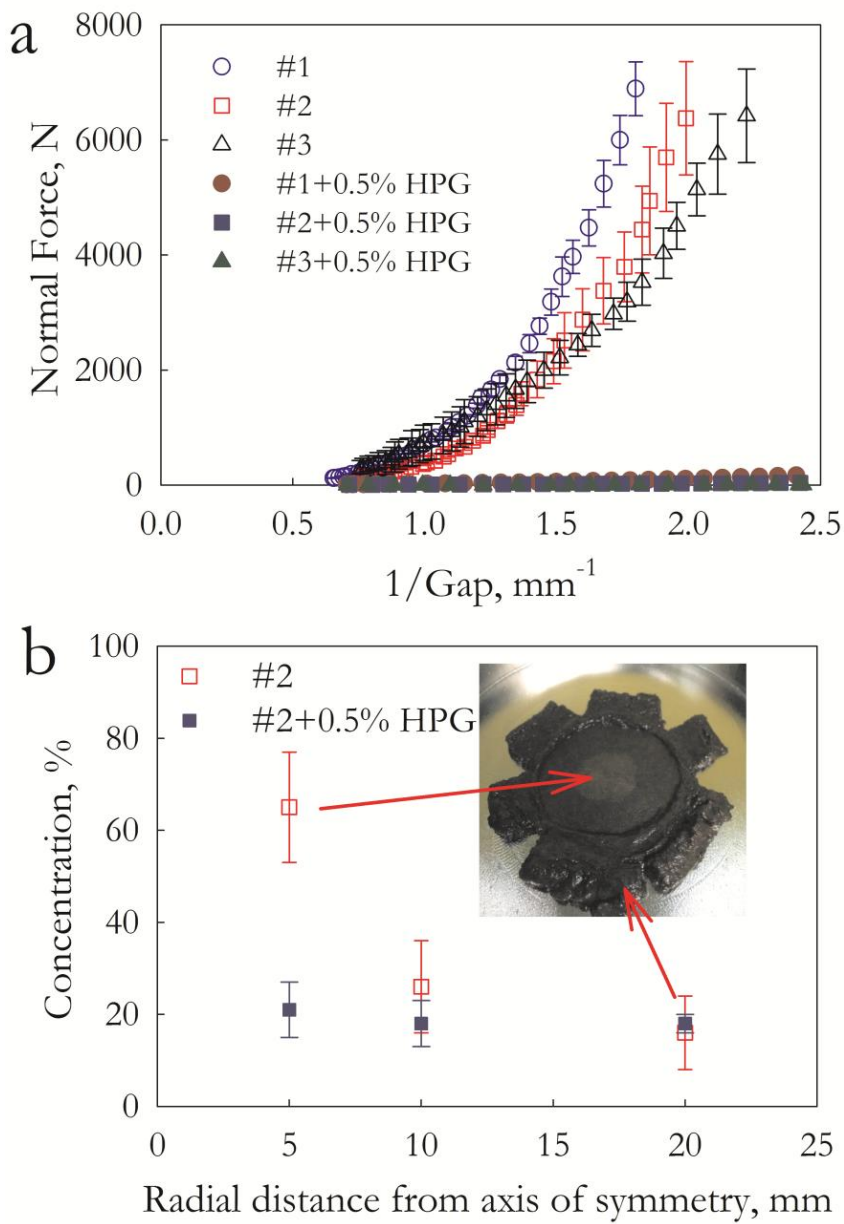

Figure 11. Normal force versus the reciprocal gap generated during squeeze flow for Suspensions #1, #2, #3 a) without the gelation agent and with 0.5 wt% of gelation agent, b) Cellulose concentration of the squeezed Suspension #2 versus radial distance from the axis of symmetry with 0.5 wt% of HPG and without HPG. The inset shows the typical image of a squeezed sample.

|  | #1 | #2 | #3 | #4 |
|---|---|---|---|---|
| Solids weight fraction | 0.209 | 0.198 | 0.25 | 0.3 |
| Solids volume fraction | 0.15 | 0.14 | 0.18 | 0.227 |
| Fiber mean diameter, μm | 12.2 | 14.3 | 14.8 | 35 |
| Mean aspect ratio | 5.9 | 6.2 | 7 | 3 |

Table 1. Weight and volume fractions of the cellulose fibers of the suspensions investigated; mean diameter and aspect ratio of the fibers of each suspension.